\documentclass[runningheads]{llncs}
\usepackage[dvipsnames]{xcolor}

\usepackage{graphicx}
\usepackage{tikz}
\usepackage{appendix}
\usepackage{amsmath}
\usepackage{amssymb}
\usepackage{bm}
\usepackage[export]{adjustbox}
\usepackage[outercaption]{sidecap}
\usepackage[labelfont=bf]{caption}
\usepackage{subcaption}
\usepackage{hyperref}

\usepackage{booktabs,array}
\newcolumntype{P}[1]{>{\centering\arraybackslash}p{#1}}

\begin{document}

\title{How Good Are Synthetic Medical Images? An Empirical Study with Lung Ultrasound} 

\titlerunning{Synthetic Images Study with Lung Ultrasound}

\author{
Menghan Yu\inst{1,2}
\and
Sourabh Kulhare\inst{1}
\and
Courosh Mehanian\inst{1}
\and
Charles B Delahunt\inst{1}
\and
Daniel E Shea\inst{1}
\and
Zohreh Laverriere\inst{1}
\and
Ishan Shah\inst{1}
\and
Matthew P Horning\inst{1}
}

\authorrunning{M. Yu, et al.}
%

\institute{
Global Health Labs, Inc, Bellevue, WA \and
University of Washington, Seattle WA 98105, USA \\
\email{(sourabh.kulhare, menghan.yu)@ghlabs.org}
}

\maketitle              

\begin{abstract}

Acquiring large quantities of data and annotations is known to be effective for developing high-performing deep learning models, but is difficult and expensive to do in the healthcare context. Adding synthetic training data using generative models offers a low-cost method to deal effectively with the data scarcity challenge, and can also address data imbalance and patient privacy issues. In this study, we propose a comprehensive framework that fits seamlessly into model development workflows for medical image analysis. We demonstrate, with datasets of varying size, (i) the benefits of generative models as a data augmentation method; (ii) how adversarial methods can protect patient privacy via data substitution; (iii) novel performance metrics for these use cases by testing models on real holdout data. We show that training with both synthetic and real data outperforms training with real data alone, and that models trained solely with synthetic data approach their real-only counterparts.
Code is available at https://github.com/Global-Health-Labs/US-DCGAN.
\end{abstract}

\section{Introduction}

The great successes of deep learning have depended on massive datasets with ground truth labels provided by domain experts. But in the healthcare context, the amount of data available for model development is often limited due to privacy, security, legal, cost, and resource availability issues. The latter two are especially challenging in low-resource settings, such as low- and middle-income countries (LMIC). Negative cases are usually easily obtainable but the paucity of positive cases can be a bottleneck, especially for rare conditions. 
\let\thefootnote\relax\footnotetext{Funding provided by Global Health Labs, Inc. (\url{www.ghlabs.org})}

Various methods have been proposed to address data scarcity. Here, we explore the use of Generative Adversarial Networks (GANs), introduced by Goodfellow \textit{et al.} \cite{goodfellow2014generative}, which generate synthetic images that closely resemble images in a particular domain. This raises the intriguing possibility of using synthetic images in a medical context to simultaneously address scarcity, class imbalance, and privacy concerns. Synthetic images with and without the feature of interest can be generated, which can both balance classes and increase the total number of samples. Furthermore, synthetic images are not traceable to any specific patient, being based on a random vector, thus maintaining patient privacy. 

Recently, a number of works \cite{nie2018medical,chuquicusma2018fool,lahiri2017generative,frid2018synthetic} have applied generative algorithms in the medical image domain. Hu et al. \cite{hu2017freehand} explored using conditional GANs to simulate fetal ultrasound images at a given 3D spatial location relative to the patient's anatomy. SpeckleGAN \cite{bargsten2020specklegan} integrates a network module with speckle noise into a GAN architecture to generate realistic intravascular ultrasound (IVUS) images. Other studies illustrate the effectiveness of GANs in generating synthetic breast ultrasound \cite{haq2023ultrasound,pang2021semi}, thyroid ultrasound \cite{liang2021data}, and transcranial focused ultrasound \cite{koh2021acoustic}, showcasing their utility in various applications. Deep Convolutional Generative Adversarial Networks (DCGAN) \cite{radford2015unsupervised} has emerged as a important building block of many generative approaches.

Despite notable advances in generating synthetic images using GANs, quantitative performance evaluation remains underdeveloped. Many of the current evaluation methods rely on subjective visual inspection, which makes it challenging in medical domains like ultrasound that require radiology expertise. Xu et al. \cite{xu2018metrics} enumerate existing quantitative metrics and discuss their strengths and limitations. While these metrics are an improvement over subjective evaluations, they do not measure the usefulness of synthetically generated images for a downstream task. In our evaluation of GAN performance, we go beyond metrics that evaluate distributional characteristics. We also assess their performance for a clinically relevant downstream task. 

To show the feasibility of this approach, we develop a comprehensive framework for synthetic image generation that fits seamlessly into model development workflows for medical image analysis. Concretely, we apply the framework to pneumonia, which remains a major global health concern. While pneumonia is no longer a problem in rich countries, it remains the leading cause of child mortality in LMICs \cite{whopna}. Accurate and timely diagnosis of pediatric pneumonia can expedite treatment and potentially save children's lives.

\begin{figure*}
     \raggedright
     \begin{subfigure}[l]{0.35\textwidth}
         \includegraphics[scale=0.125]{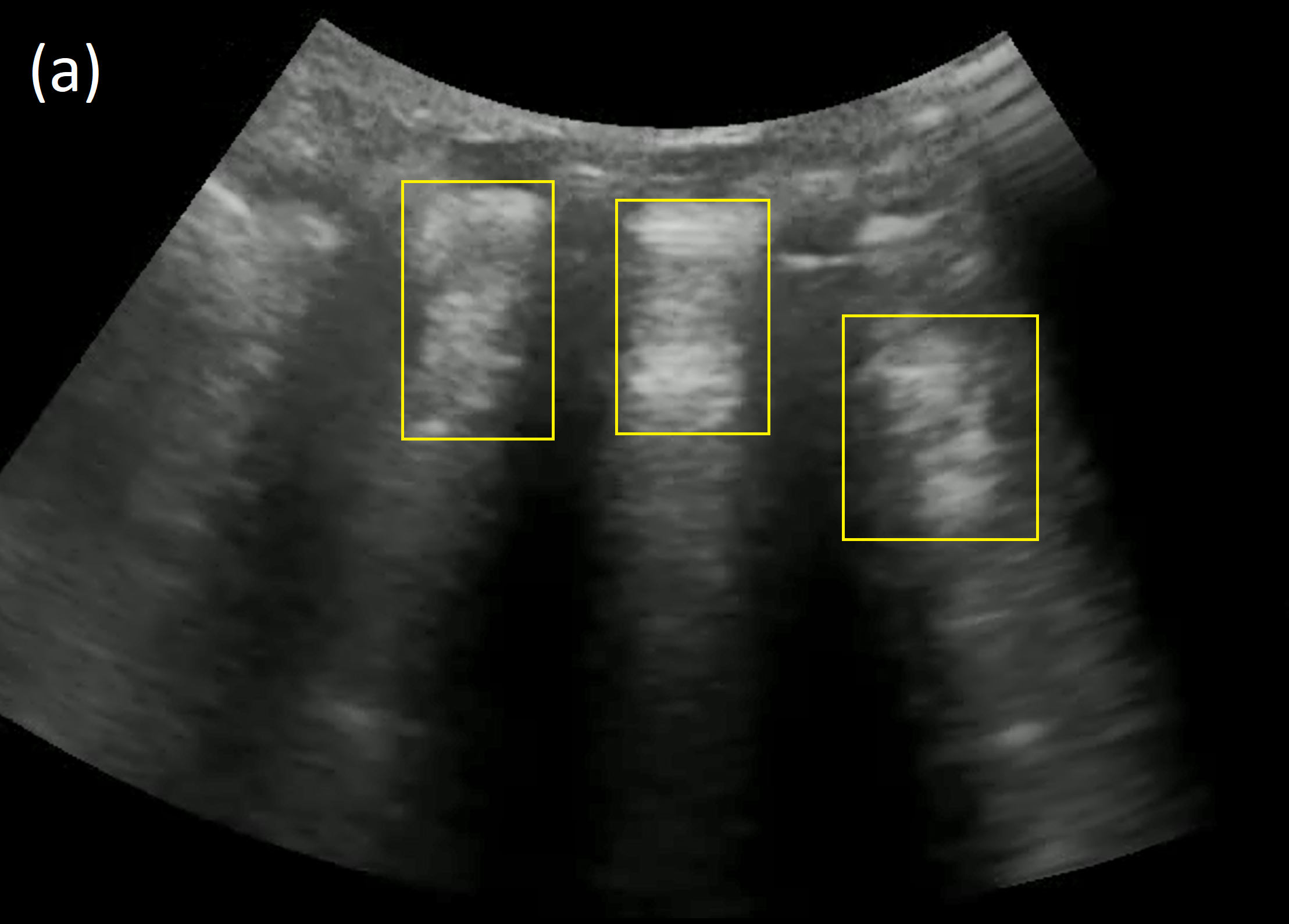}
         \label{consolidation:real}
     \end{subfigure}
     \hspace{16mm}
     \begin{subfigure}[r]{0.35\textwidth}
         \includegraphics[scale=0.125]{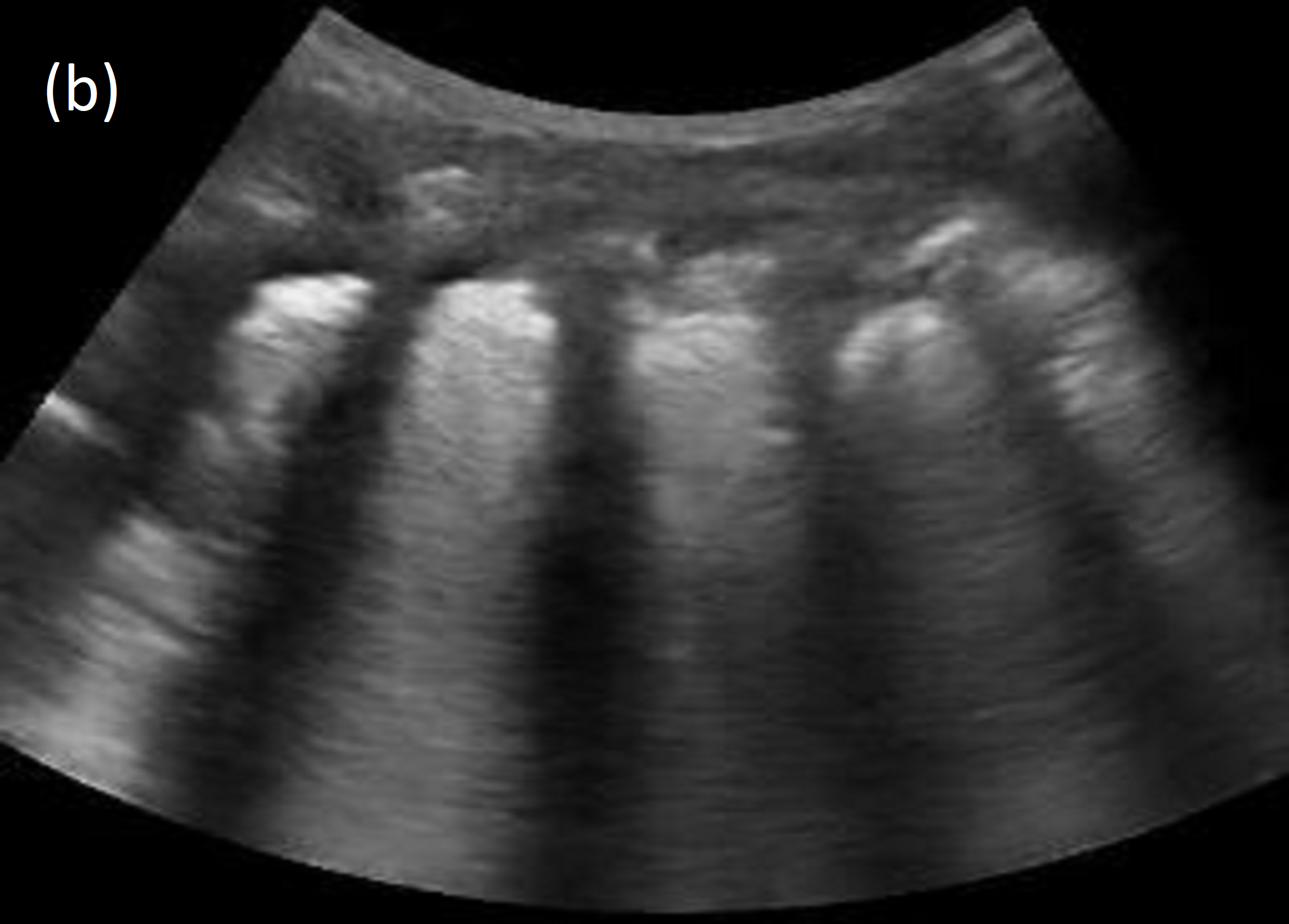}
         \label{consolidation:synthetic}
     \end{subfigure}
     \vspace{-2mm}
     \caption{Consolidation in lung ultrasound. (a) Real consolidation image: infection-induced inflammation causes fluid buildup in the alveoli, presenting as hyper- and hypo-echoic splotches (yellow boxes). (b) Synthetic consolidation image.}
     \label{consolidation}
\end{figure*}

The standard-of-care for confirmatory imaging in high-income countries is X-ray, which is often not an option in LMICs. Lung ultrasound has emerged as a promising alternative with many advantages including safety, portability, and low cost. However, confirming pneumonia with lung ultrasound is challenging because images mainly consist of acoustic artifacts whose interpretation requires expertise, which is lacking in LMICs.  Deep learning has proved to be an effective tool to assist healthcare workers to identify lung ultrasound features associated with pneumonia \cite{kulhare2018,ouyang2023,shea2023}, the primary one being consolidation (see Figure\,\ref{consolidation}).

Our contributions are as follows:
\vspace{-2mm}
\begin{itemize}
  \item We apply DCGAN to generate synthetic images for a medically high-impact downstream task: consolidation classification. To the best of our knowledge, our work represents the first application of DCGAN to lung ultrasound.
  \item We present novel quantitative metrics that measure real and synthetic image distribution similarity and the presence of image features of interest.
  \item We show the benefit of synthetic data in two crucial use cases: augmentation of scarce data, and protection of patient privacy via data substitution.
\end{itemize}

\vspace{-6mm}

\section{Methods} 

\subsection{Dataset}

\subsubsection{Description and Collection Process.}
A study conducted from 2017-2022 in Nigeria collected lung ultrasound data at three sites comprising $550$ pediatric  patients (\textit{i.e.}, $< 18$ years old, $29\%$ female, $71\%$ male). The data has been described previously \cite{shea2023}. Briefly, ultrasound cine loops were collected (with the Mindray DP-10 with 35C50EB curvilinear probe) from 10 lung zones covering anterior, lateral, and posterior locations consistent with the international lung ultrasound protocol \cite{international}. Radiology (X-ray) of the lungs was performed prior to the ultrasound exam and, combined with clinical indications, was used to screen patients and identify ground truth diagnosis. Table \ref{sample-table} lists patient, video, and frame counts across training, validation, and holdout sets. Additional details on device settings, data collection protocol, and patient demographics are provided in the supplement.

\begin{table*}[ht!]
  \setlength{\tabcolsep}{2.5mm}
  \centering
  \caption{Counts of negative and positive patients, videos, and frames, in training, validation, and testing sets, which were defined at the patient level.}
  \vspace{3mm}
  \begin{tabular}{ l r r | r r | r r }
    Pediatric &  & Training &  &   Validation &  &  Testing \\
  \toprule
    & neg & pos & neg & pos  & neg & pos \\
  \midrule
    Patients & 122 & 134 & 23 & 96 & 24 & 100  \\
    Videos & 5,022 & 1,134 & 1,229 & 404 & 695 & 423  \\
    Frames & 426,675 & 99,134 & 163,705 & 58,580 & 117,391 & 64,955  \\
  \bottomrule
  \end{tabular}
  \label{sample-table}
\end{table*}

\subsubsection{Annotation process.}
Ultrasound data were annotated at both the video and frame levels using a custom web-based tool. Annotations were performed by expert lung ultrasound physicians to identify the following lung features: pleural line, A-line, B-line, merged B-line, abnormal pleural line, pleural effusion, and consolidation \cite{bhoil2021signs}. For this work, videos with consolidation were used as feature positive videos and videos with no consolidation were used as negative videos. Frame annotations were used as ground truth for feature-positive frames to train and evaluate CNN classifiers.

\subsubsection{Preprocessing.}
Ultrasound videos comprised 145 frames on average. Frames were centrally cropped using fixed cropping points and masked to remove extraneous text. Frames were resized to $256 \times 256$ and pixel values standardized to zero mean and unit standard deviation for both image generation and downstream classification tasks. To ensure clean frame labels, positive frames were selected
only from consolidation-positive videos, and negative frames from consolidation-negative videos.

\subsection{Deep Convolutional Generative Adversarial Networks}

We use a backbone network similar to the Deep Convolutional Generative Adversarial Networks (DCGAN) architecture \cite{radford2015unsupervised}, since it addresses the instability of vanilla GANs and allows arbitrary output image resolution. The input tensor size of the discriminator and the output of the generator were set to $256 \times 256 \times 1$, which is fine enough to capture lung ultrasound details and enables the use of off-the-shelf pre-trained models \cite{krizhevsky2017imagenet} to evaluate synthetic images. The generator consists of a cascade of strided 2D transposed convolutional layers, batch normalizations, and ReLu activations, with an input latent vector $\bm{z}$ that is drawn from a standard normal distribution. The discriminator comprises a cascade of 2D strided convolutional layers, batch normalization layers, and leaky ReLu activations. We stabilized the adversarial training process by using a smaller batch size of 16, lowering generator and discriminator learning rates ($1 \times 10^{-5}$ and $5 \times 10^{-6}$ respectively), and adding a dropout layer (rate 0.25) after each convolutional block of the discriminator. Full details can be found in the code.

\subsection{Evaluations}

\subsubsection{Qualitative metrics.}
Synthesized images are assessed for the presence of features commonly observed in lung ultrasound. Details are described in Section\,\ref{results_and_discussion}.

\subsubsection{Quantitative metrics.}
\label{quantmetrics}
We adopt three metrics to (i) monitor GAN convergence during training, (ii) choose the best GAN model, and (iii) evaluate image quality. Following \cite{xu2018metrics}, we use (a) kernel Maximum Mean Discrepancy (kMMD), which measures the distance between the real and synthetic image distributions. At every GAN training epoch, we randomly sample $2,000$ images from the training set and use the current GAN model to randomly generate $2,000$ synthetic images. We estimate kMMD empirically over image pairs from both sets. If real and synthetic distributions closely match, mean kMMD should be near $0$. 

Also following \cite{xu2018metrics}, we use (b) a 1-Nearest Neighbor (1NN) classifier to distinguish real and synthetic positive images, which is trained and evaluated in a leave-one-out (LOO) fashion. Zero accuracy indicates overfitting to the real image distribution, while unit accuracy means perfect separability, indicating non-overlap between the real and generated distributions. Ideally, 1NN accuracy should be near $0.5$ (\textit{i.e.}, chance). For (a) and (b), images are represented by the feature embedding generated by an ImageNet pre-trained ResNet34 model to ensure that meaningful image structural information is captured. Feature space image representation is described in more detail in the supplement.

\begin{figure}[hbt!]
\includegraphics[scale=0.62]{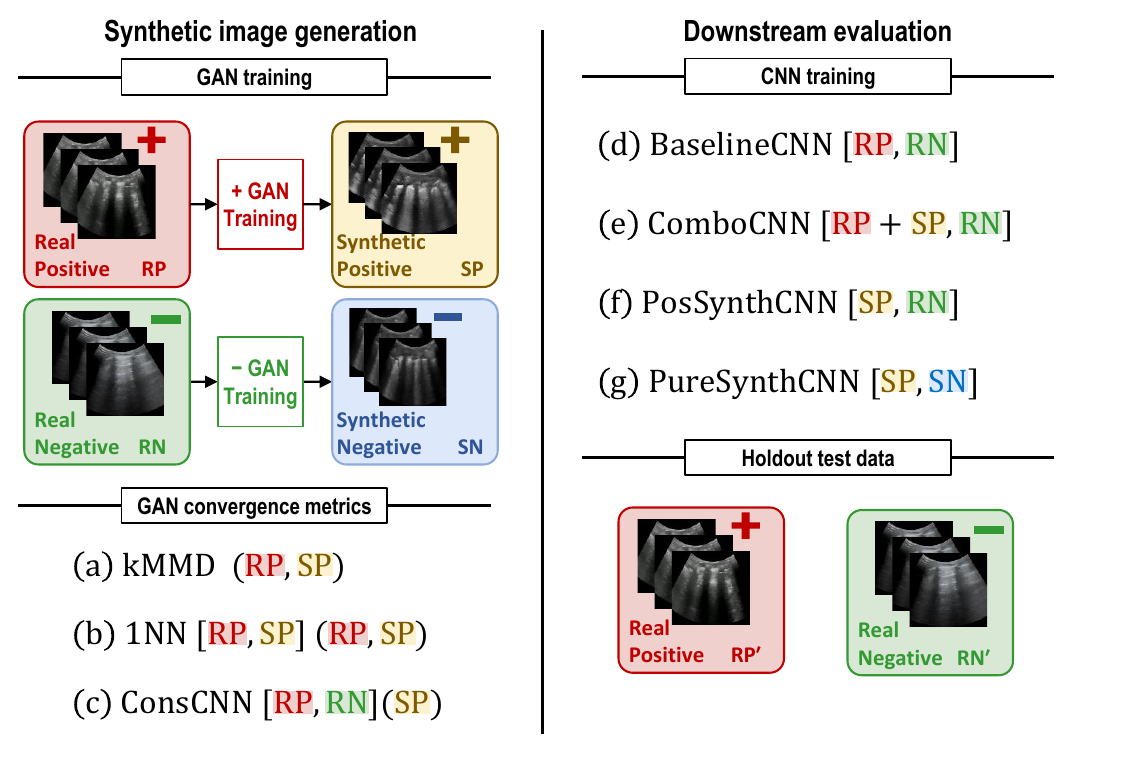} 
\vspace{-2mm}
\caption{GAN training and evaluation workflow. (Top-left) Positive and negative GANs are trained with real positive and real negative images, respectively. (Bottom-left) Quantitative metrics used to monitor GAN convergence during training (Figure\,\ref{confidence}). Square brackets [,] indicate training sets; parentheses (,) testing sets. (a) kMMD (no training) measures distance between real and synthetic distributions. (b) 1-NN is trained to distinguish real and synthetic images and is tested in leave-one-out-fashion. (c) Consolidation CNN is trained to recognize consolidation and is tested on synthetic positive images. (Top-right) Downstream evaluations measure GAN effectiveness by training various CNNs and testing on real holdout data (Figure\,\ref{auroc}). (d) Baseline CNN is trained on real positive and negative images. (e) Combined CNN is trained on real and synthetic positive images and real negative images. (f) Positive synthetic CNN is trained on synthetic positive and real negative images. (g) Pure synthetic CNN is trained on synthetic positive and negative images (results in Table\,\ref{fullsynthetic}). (Bottom-right) Holdout set of real positive and negative images is used for downstream CNN testing.}
\label{process_new}
\end{figure}

Finally, we use (c) a CNN consolidation classifier, which is trained in advance on all real samples, to assess the presence of consolidation in the synthetic positive images. Average confidence scores near unity indicate consolidation features are present in the images. The workflow for GAN training and evaluation is shown in Figure\,\ref{process_new}. Convergence metric (c) is not used for negative GAN training.

\subsubsection{Downstream Task Evaluations.} \label{downstream}
We envision two scenarios: synthetic data can either augment or entirely replace training data. To evaluate the effectiveness of synthetic data in a realistic downstream task, we train CNNs for binary consolidation classification on a frame-by-frame basis and assess frame-level performance as a function of the amount of real training data used. The output of these consolidation CNNs is a confidence score and the relevant performance metric is the AUC-ROC performance on real hold-out test data.

Four classifiers with identical CNN architectures (reduced VGG \cite{simonyan2015deep}) are trained to recognize consolidation, with various positive and negative training sets as follows: Baseline classifier (d) is trained with real positive and negative frames. Combined classifier (e) uses both synthetic and real positive frames and real negative frames. Positive synthetic classifier (f) uses synthetic positive frames and real negative frames. Pure synthetic classifier (g) uses only synthetic images for both positive and negative sets. All classifiers were trained under consistent ``best known" practices: standard data augmentation techniques applied to training data, including horizontal flipping, gamma brightness adjustment, Gaussian blurring, and random pixel intensity adjustment. We also did extensive hyper-parameter optimization (using random search to tune learning rate, channel multiplier, and dropout rate). Only the model with the highest validation accuracy is selected for evaluation on the holdout set.

\section{Results}
\label{results_and_discussion}

\subsection{Qualitative evaluation}

\begin{figure}[ht!]
  \centering
  \includegraphics[width=12.5cm]{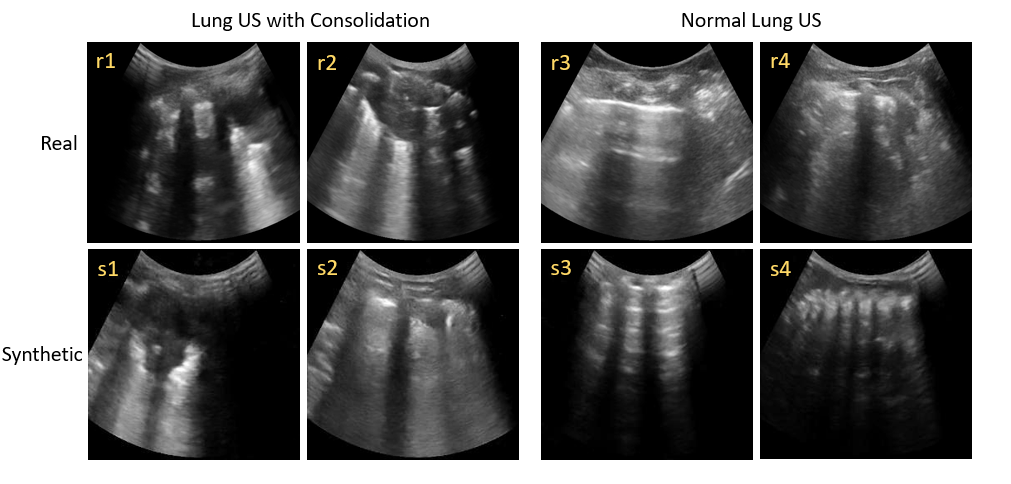}
  \caption{Examples of real and synthetic images.}
  \label{realandsynthetic}
\end{figure}

Lung ultrasound images largely consist of artifacts generated by acoustic impedance mismatches between lung tissue, fluid, and air. To the trained eye, these artifacts convey information about lung pathologies such as consolidation, which manifests as sub-pleural hyper- and hypo-echoic splotches caused by inflammatory alveolar fluid buildup. See Figure\,\ref{consolidation}. We observe that the synthetic images also display such features and artifacts and preserve the architecture of lung ultrasound images. Figure\,\ref{realandsynthetic} shows examples of synthetic and real, consolidated and normal, lung ultrasound images. Prominent features of lung ultrasound include the ``bat sign" \cite{bhoil2021signs} and ``rib shadows", caused by the pleural line and the impenetrability of ribs to ultrasound. These can be observed in the real and synthetic lung images (r1--r4, s1--s4). Pleural lines, A-lines (multiple-echo artifacts of the pleural line) are also present and qualitatively similar in synthetic images (r3, s3). The ``shadows" on the left and right sides of images (r1, r2, r4, s1--s4) are caused by poor acoustic contact between the edges of the curvilinear transducer and the child's body, which is another realistic element of the the synthetic images. Other features of lung ultrasound also appear in synthetic images, such as ``hepatization", caused by excessive fluid accumulation, as seen in (s1, s2). Additional synthetic examples are provided in the supplement.

\vspace{-2mm}
\subsection{Quantitative evaluation}

\begin{figure}[hbt!]
  \begin{subfigure}[l]{0.325\textwidth}
    \includegraphics[scale=0.3]{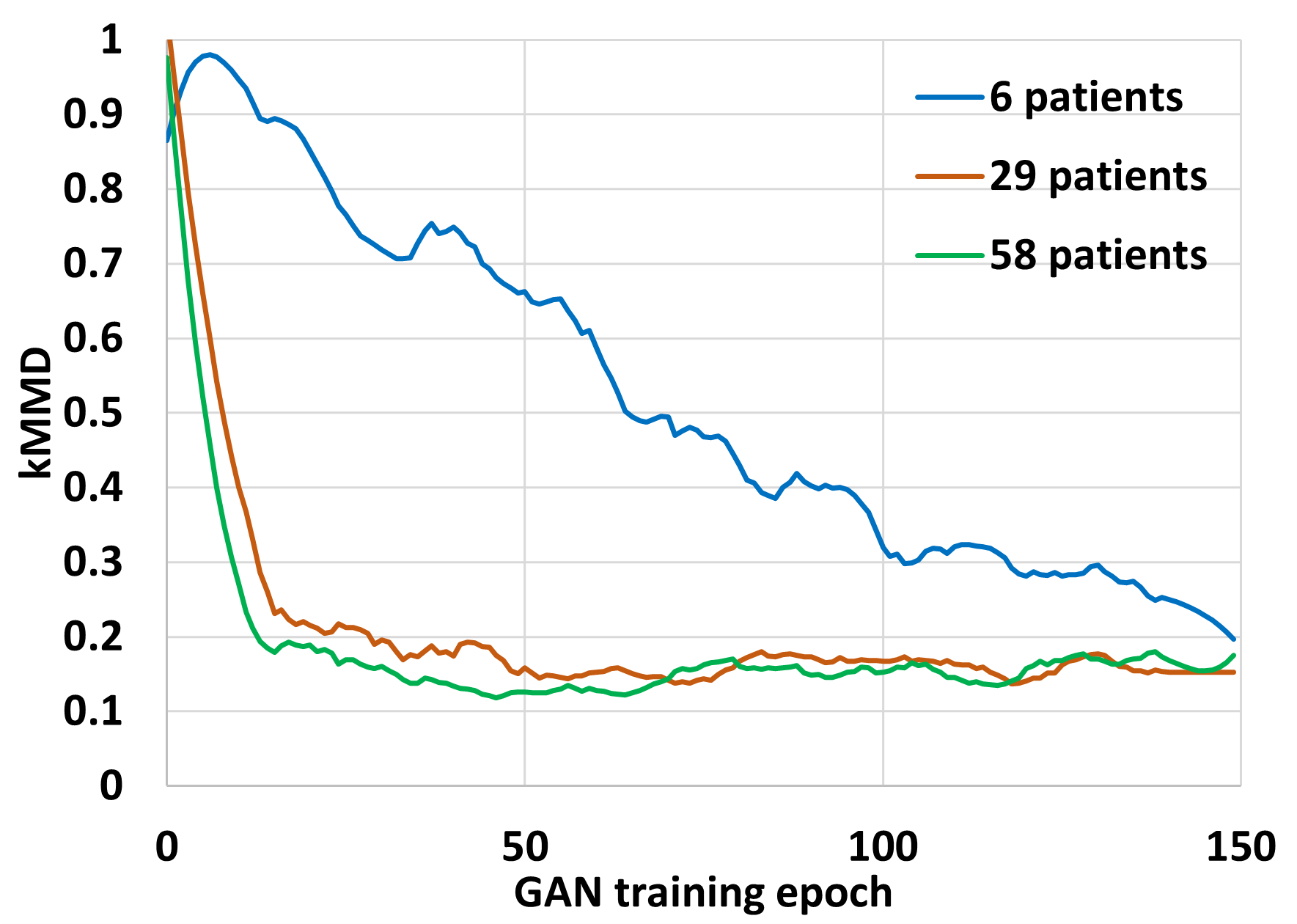}
  \end{subfigure}
  \begin{subfigure}[r]{0.325\textwidth}
    \includegraphics[scale=0.3]{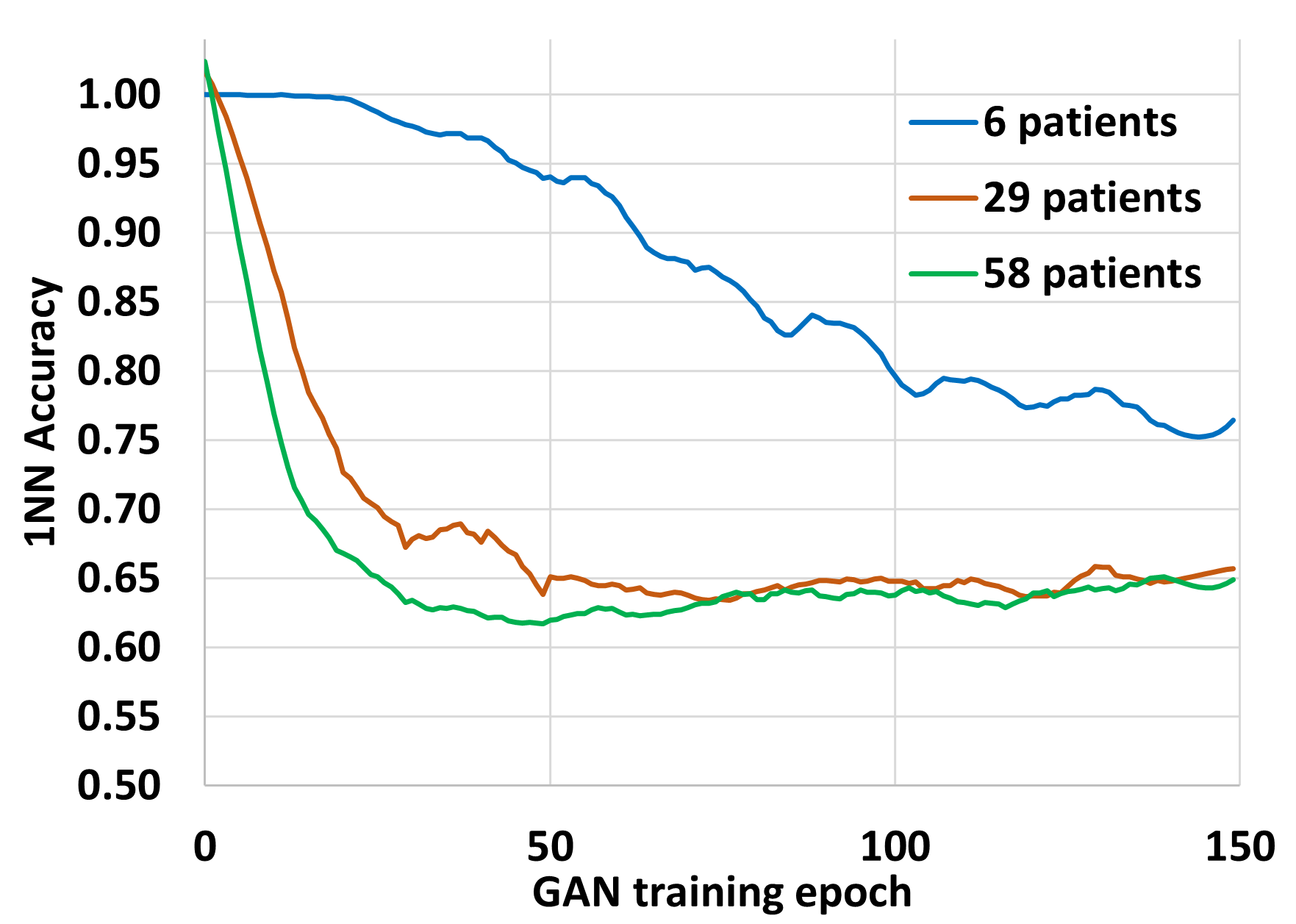}
  \end{subfigure}
  \begin{subfigure}[r]{0.325\textwidth}
    \includegraphics[scale=0.3]{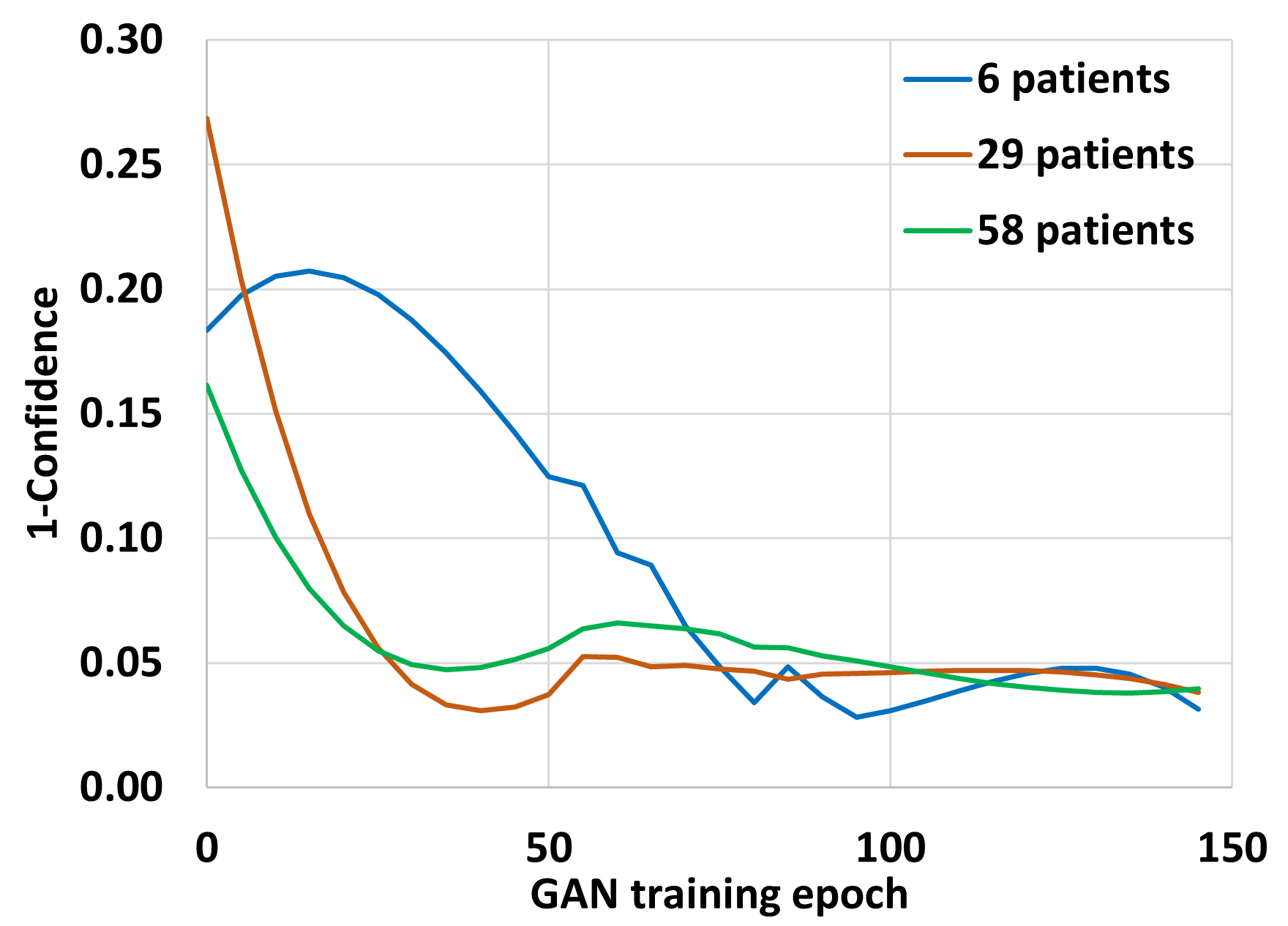}
  \end{subfigure}
  \caption{Quantitative metrics for positive GAN training vs epoch for different dataset sizes. Left: kMMD. Center: 1NN accuracy. Right: $1-$ confidence score. Blue, orange, and green represent 6, 29, and 58 patient training sets, respectively.}
  \label{confidence}  
\end{figure}

Figure \ref{confidence} plots quantitative evaluation metrics (kMMD, 1NN LOO accuracy, and $1-$ mean consolidation score as described in Section\,\ref{quantmetrics}) for different training set sizes as a function of GAN training epoch. All metrics decreased over the course of training as expected and plateaued near 60 epochs. For the consolidation classifier, the mean score reached 0.96 for all training set sizes, indicating the success of synthesizing the consolidation feature. We note that kMMD and 1NN LOO accuracy did not reach their ideal levels as defined in Section\,\ref{quantmetrics}.

\subsection{Downstream evaluation}

\begin{figure}[hbt!]
\includegraphics[width=\textwidth]{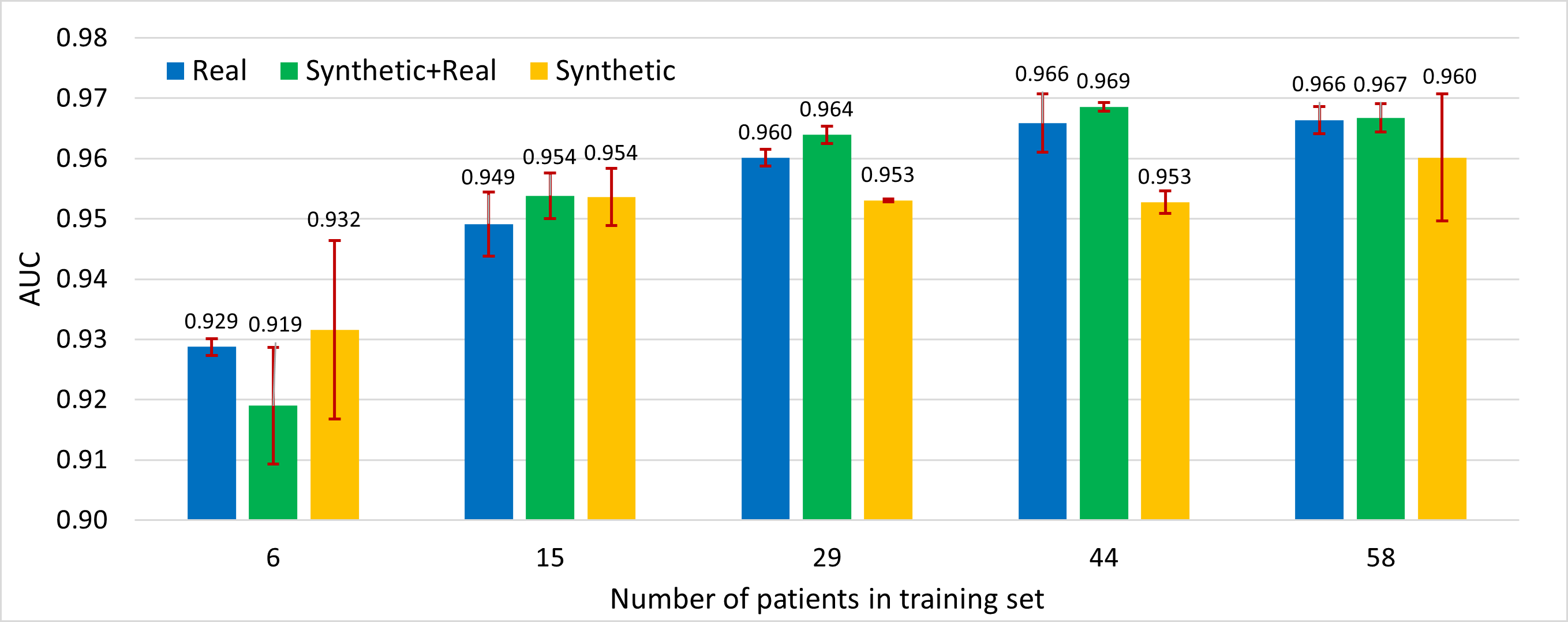}
\caption{Downstream evaluation results on real holdout test set for models trained as follows: blue: real only (baseline); green: real augmented with synthetic positive images; orange: synthetic positive images only. Green improvement over blue shows the benefit of synthetic augmentation. Approximate equivalence between orange and blue shows that synthetic data is a good proxy for real data.} 
\label{auroc}
\end{figure}

Figure \ref{auroc} shows downstream results for different amounts of training data (patient counts of 6, 15, 29, 44, and 58). AUC-ROC on a holdout set containing only real images is plotted. Balanced positive and negative frame counts were maintained during the patient sampling process. The figure displays three bars for each dataset size: (d) blue for the baseline model trained solely on real data, (e) green for the model trained by augmenting positives with synthetic data, and (f) yellow for the model trained on synthetic data as proxy for positive data (as described in Section\,\ref{downstream}). To ensure robustness, each data ablation point was averaged over three random patient sampling replicates. 

Our results highlight an advantage in training models by combining real patient data with GAN-generated synthetic data (except with very sparse training data). This can be seen as the improvement between the green and blue bars in Figure\,\ref{auroc} for training dataset size $\geq 15$ patients, where the AUC-ROC averaged over replicates increased from 0.949 to 0.954. At this dataset size and greater, the GAN can generate high-quality consolidation ultrasound images that effectively cover the real data distribution.  

Additionally, training models solely on synthetic positive images achieved comparable accuracy to models trained on real data only, as evidenced by the rough equivalence between the blue and orange bars in Figure\,\ref{auroc}. This finding reduces the required amount of data to be collected and shared to develop high-performing models, thus mitigating privacy concerns and development costs.

Our work has primarily focused on generating feature-positive samples, which are often harder to obtain in a healthcare context. However, synthetic feature-negative samples are generated the same way, and are also useful for the patient privacy use case. This motivated the development of downstream task (g) described in Section\,\ref{downstream}. For this model, the training set consists entirely of synthetic data, both positive and negative. Table \ref{fullsynthetic} presents the classifier results on the same holdout set as for the other downstream tasks. 

\vspace{-2mm}
\begin{table*}
  \setlength{\tabcolsep}{4mm}
  \centering
  \caption{Positive and negative GANs were trained with real positive and negative frames, respectively. Binary classifiers were trained using 67,950 generated frames for each class at various GAN training epochs. Test results are shown.}
  \vspace{4mm}
  \begin{tabular}{lcccc}
    Epoch & 25 & 50 & 75 & 100 \\
    \midrule
    Accuracy & 0.903 & 0.912 & 0.916 & 0.922 \\
    AUCROC & 0.952 & 0.960 & 0.954 & 0.962 \\
  \bottomrule
  \end{tabular}
  \vspace{3mm}
  \label{fullsynthetic}
\end{table*} 

\section{Discussion}

Deep learning has made tremendous strides in robust classification of medical images over the last decade and high-performing models are critical to using AI technology to address resource shortages in LMICs. But deep learning requires massive amounts of data to train clinically viable models. Generative adversarial networks have proved to be a useful tool in offsetting the amount of data required to train these models, thereby reducing cost and amplifying the impact of modest investments. At the same time, synthesized images can help to mitigate concern over patient privacy and balance data for rare conditions. These benefits improve the prospects for developing robust and unbiased models that can make a positive impact in healthcare in LMICs.

We have presented a framework for generating synthetic images and demonstrated its effectiveness using the healthcare-relevant example of consolidation in lung ultrasound images. Through a comprehensive suite of qualitative and quantitative assessment metrics, we have shown the model's proficiency in capturing the target distribution and the presence of the desired image features. Crucially, we have shown that data augmentation with synthetic images bolsters the performance of clinically relevant downstream tasks, which is the ultimate test of the usefulness of synthetic images generated with GANs. Future work will incorporate image representations based on ultrasound-specific embeddings and feedback from radiologists on synthetic medical image quality.


%
\bibliography{mybibliography}{}

\begin{thebibliography}{10}
\providecommand{\url}[1]{\texttt{#1}}
\providecommand{\urlprefix}{URL }
\providecommand{\doi}[1]{https://doi.org/#1}

\bibitem{bargsten2020specklegan}
Bargsten, L., Schlaefer, A.: Speckle{GAN}: a generative adversarial network with an adaptive speckle layer to augment limited training data for ultrasound image processing. IJCARS  \textbf{15},  1427--1436 (2020)

\bibitem{bhoil2021signs}
Bhoil, R., Ahluwalia, A., Chopra, R., Surya, M., Bhoil, S.: Signs and lines in lung ultrasound. J. Ultrason.  \textbf{21}(86),  225--233 (2021)

\bibitem{chuquicusma2018fool}
Chuquicusma, M.J., et~al.: How to fool radiologists with {GANs}? {A} visual {Turing} test for lung cancer diagnosis. In: ISBI. pp. 240--244. IEEE (2018)

\bibitem{frid2018synthetic}
Frid-Adar, M., et~al.: Synthetic data augmentation using gan for improved liver lesion classification. In: ISBI. pp. 289--293. IEEE (2018)

\bibitem{goodfellow2014generative}
Goodfellow, I., et~al.: Generative adversarial networks. NeurIPS  (2014)

\bibitem{haq2023ultrasound}
Haq, D.Z., Fatichah, C.: Ultrasound image synthetic generating using deep convolution generative adversarial network for breast cancer identification. IPTEK The Journal for Technology and Science  \textbf{34}(1), ~12 (2023)

\bibitem{hu2017freehand}
Hu, Y., et~al.: Freehand ultrasound image simulation with spatially-conditioned {GANs}. In: Molecular Imaging, Reconstruction and Analysis of Moving Body Organs, and Stroke Imaging and Treatment. pp. 105--115. Springer (2017)

\bibitem{koh2021acoustic}
Koh, H., et~al.: Acoustic simulation for transcranial focused ultrasound using gan-based synthetic ct. IEEE J. Biomed. Health Inform.  \textbf{26}(1),  161--171 (2021)

\bibitem{krizhevsky2017imagenet}
Krizhevsky, A., Sutskever, I., Hinton, G.E.: Imagenet classification with deep convolutional neural networks. Communications of the ACM  \textbf{60}(6),  84--90 (2017)

\bibitem{kulhare2018}
Kulhare, S., et~al.: Ultrasound-based detection of lung abnormalities using single shot detection convolutional neural networks. In: Simulation, image processing, and ultrasound systems for assisted diagnosis and navigation. pp. 65--73 (2018)

\bibitem{lahiri2017generative}
Lahiri, A., Ayush, K., Kumar~Biswas, P., Mitra, P.: Generative adversarial learning for reducing manual annotation in semantic segmentation on large scale miscroscopy images: Automated vessel segmentation in retinal fundus image as test case. In: Proceedings of CVPR Workshops. pp. 42--48 (2017)

\bibitem{liang2021data}
Liang, J., Chen, J.: Data augmentation of thyroid ultrasound images using generative adversarial network. In: IEEE IUS. pp.~1--4. IEEE (2021)

\bibitem{nie2018medical}
Nie, D., et~al.: Medical image synthesis with deep convolutional adversarial networks. IEEE TBME  \textbf{65}(12),  2720--2730 (2018)

\bibitem{ouyang2023}
Ouyang, J., et~al.: Weakly semi-supervised detection in lung ultrasound videos. In: IPMI 2023. pp. 195--207 (2023)

\bibitem{pang2021semi}
Pang, T., Wong, J.H.D., Ng, W.L., Chan, C.S.: Semi-supervised gan-based radiomics model for data augmentation in breast ultrasound mass classification. Computer Methods and Programs in Biomedicine  \textbf{203},  106018 (2021)

\bibitem{radford2015unsupervised}
Radford, A., et~al.: Unsupervised representation learning with deep convolutional generative adversarial networks. arXiv preprint arXiv:1511.06434  (2015)

\bibitem{shea2023}
Shea, D., et~al.: Deep learning video classification of lung ultrasound features associated with pneumonia. In: CVPR 2023. pp. 3102--3111. IEEE (2023)

\bibitem{simonyan2015deep}
Simonyan, K., Zisserman, A.: Very deep convolutional networks for large-scale image recognition. arXiv preprint arXiv:1409.1556  (2015)

\bibitem{international}
Volpicelli, G., et~al.: International evidence-based recommendations for point-of-care lung ultrasound. Int. J. Med. Inform.  \textbf{129},  413--422 (2019)

\bibitem{whopna}
{W}orld {H}ealth~{O}rganization ({WHO}): Pneumonia (2016), \url{http://www.who.int/mediacentre/factsheets/fs331/en}

\bibitem{xu2018metrics}
Xu, Q., el~al.: An empirical study on evaluation metrics of generative adversarial networks. arXiv preprint arXiv:1806.07755  (2018)

\end{thebibliography}
\bibliographystyle{splncs04}

\end{document}